%% file: main.tex
\documentclass[10pt,journal]{IEEEtran}

\usepackage{cite}
\usepackage{amsmath,amssymb,amsfonts}
\usepackage{graphicx}
\usepackage{textcomp}
\usepackage{xcolor}
\usepackage{booktabs}
\usepackage{multirow}
\usepackage{balance}
\usepackage[hidelinks]{hyperref}

\newcommand{\parallaxgitrepo}{\url{https://github.com/karusankaralingam/PARALLAX}}

\newif\ifappendix
\appendixfalse

\begin{document}

\title{Can LLMs Perform Technical Comprehension of Computer Architecture Papers?}

\author{Nishant Aggarwal, Aishwarya Lekshmi Chithra, Ayushi Dubal, Sreeraj Kannakarankodi, Ian McDougall, Adarsh Mittal, Vishnu Ramadas, Noah Scott, Ranganath Selagamsetty, Weichu Yang, and Karthikeyan Sankaralingam%
\thanks{University of Wisconsin--Madison and NVIDIA Research.}}

\maketitle

\begin{abstract}
\input{abstract}
\end{abstract}


\input{sec_introduction}
\input{sec_related}
\input{sec_gauntlet}
\input{sec_expt_design}
\input{sec_results}
\input{sec_discussion}

\balance
\bibliographystyle{IEEEtran}
\bibliography{references}

\ifappendix
\input{sec_appendix}
\fi

\end{document}

%% file: abstract.tex
Can large language models perform \emph{technical comprehension} of computer architecture papers---not summarization, but structured critique that names the core mechanism, surfaces buried assumptions, and connects a contribution beyond its own scope?
We study Gauntlet, an open-source pipeline that analyzes a paper through five independent expert-persona reviewers and an adversarial synthesis stage.
On 20 ISCA 2025 and HPCA 2026 papers, 10 researchers each wrote their own analyses and then judged, for papers other than their own, the human analysis against Gauntlet's.
Across the 20 comparisons evaluators preferred Gauntlet in 15 (human in 4, one tie); its advantage is significant on per-analyst totals (two-sided Wilcoxon, $p<0.001$) and largest on Critical Rigor.
Where humans win, it is on trust and usefulness rather than depth: a confident wrong claim, a mechanism described but not taught, or unprioritized breadth.
A 98-paper automated ablation shows the gain comes from the multi-agent structure: the pipeline beats the same model run as a single rich-persona agent on 96\% of papers.
We release all analyses, scores, and the rubric as a community resource.

%% file: sec_introduction.tex
\section{Introduction}
\label{sec:intro}

Keeping pace with the computer architecture literature is increasingly hard. ISCA, MICRO, and HPCA 2025 alone added more than a hundred papers across near-memory processing, accelerators, coherence, security, and ML compilation. Authors often compare against favorable baselines and leave key assumptions implicit, so a paper's real contribution can be hard to extract. Readers have little help beyond summarization, which condenses a paper without teaching it. Understanding a paper well enough to critique, build on, or teach it requires naming four things: the precise structures it builds, the non-obvious insight that makes the mechanism work, specific weaknesses within the paper, and the paper's assumptions.
We call this \emph{technical comprehension} and ask: \emph{can large language models perform it, at a level comparable to trained human researchers?} We measure this through human judgment of LLM-written analyses of computer architecture research papers, which captures how a trained reader rates an analysis rather than independently verifying each technical claim. 

\textbf{Approach.}
Simply asking a frontier model to ``comprehend this paper'' does not get you there. A single-shot analysis reads well but misses the mechanism, as our ablation shows (Section~\ref{sec:results-ablation}). Two ideas close the gap, and they are the core of this paper. First, several expert perspectives read a paper better than one; each catches concerns the others miss. Second, those perspectives are formed independently and then combined by a synthesis step that preserves their disagreements rather than averaging them away.
We implement both in our tool \href{https://github.com/VerticalResearchGroup/Gauntlet}{Gauntlet} as \emph{multi-perspective independent review followed by adversarial synthesis}. Five reviewer agents analyze the paper independently (a microarchitecture specialist, a workload and evaluation analyst, a simulation-tools auditor, and two domain specialists matched to the paper's sub-topics from a $\sim$90-persona library), and a synthesizer then integrates them (Section~\ref{sec:pipeline}). All analyses use Claude Opus 4.5.

\textbf{Framing.}
We do not claim reading a paper is unnecessary. The learning that reading produces has no substitute.
The claim is narrower: a multi-perspective pipeline produces a first-pass analysis that trained readers rate above their own on specificity and rigor, and which is good enough to take seriously for triage, crossing into unfamiliar subdomains, or as a teaching aid, provided that its technical claims are checked, since our own data shows that it can be confidently wrong.

\textbf{Is this architecture research?}
One might read this as an ML paper, or an LLM-usage paper. The leverage, though, is not generic prompting but encoded \emph{domain knowledge}. Only an architect knows to instantiate a reviewer that stress-tests benchmark selection, an auditor that probes simulator fidelity, and topic-matched specialists. The ablation bears this out: a carefully engineered single-agent prompt still loses on 96\% of 98 papers (Section~\ref{sec:results-ablation}). Applying generative AI to architecture tasks is an active line of work~\cite{alphazeromoment,archagent}, and in the tradition of CAL tools papers~\cite{10.1109/LCA.2025.3559738} this is an instrument for reading the field's own literature.

%% file: sec_related.tex
\section{Related Work}
\label{sec:related}

\textbf{LLM feedback on scientific papers.}
\cite{liang2024feedback} found GPT-4 review comments overlap human reviews by 30--39\%, comparable to the 28--35\% overlap between two humans. \cite{thakkar2025iclr}, in a 20{,}000-review randomized study at ICLR 2025, showed that reviewers given LLM feedback wrote longer, more informative reviews. A parallel line estimates that 6.5--16.9\% of recent ML review text was substantially LLM-modified~\cite{liang2024monitoring}.
These establish that LLM feedback is useful and already pervasive.
We ask a narrower question: whether an LLM can perform \emph{deep technical comprehension} of mechanism papers. We compare multi-agent against single-agent designs, rather than measuring LLM assistance to human reviewers.

\textbf{Benchmarking architectural knowledge.}
QuArch~\cite{quarch} measures whether models grasp the field's concepts through expert-curated question answering; we examine the complementary axis of open-ended comprehension of individual papers, the mechanism reconstruction and critique a researcher performs while reading.

\textbf{Evaluating generated analyses.}
Judging open-ended generated text on its own is a studied problem. Surveys of evaluation practice~\cite{van2019best} find wide variation and recommend separately defined criteria over a single quality score and explicit reporting of evaluator numbers and expertise, which we follow. For the statistical comparison we follow the convention for comparing two methods across a set of problems laid out in \cite{demvsar2006statistical}, in which the problem itself is the unit of analysis, the Wilcoxon test is preferred for small samples of non-commensurable scores, and per-dimension comparisons require an explicit correction for multiple testing.

\textbf{Multi-agent and multi-persona prompting.}
Multi-agent debate improves factuality and reasoning by having instances critique one another over rounds~\cite{du2023debate}, while Solo Performance Prompting has a single model simulate multiple personas~\cite{wang2024spp}. Both let agents influence each other during generation.
The closest precedent is MARG~\cite{darcy2024marg}, which splits a paper's \emph{sections} across agents to beat context limits and cuts generic comments from 60\% to 29\% versus a single-agent GPT-4 baseline.
We instead distribute \emph{perspectives}: each reviewer reads the whole paper from a distinct expert viewpoint, fully independently. We then add an explicitly adversarial synthesis stage that preserves disagreement rather than merging it.

\textbf{LLM-as-judge.}
Our ablation uses an LLM judge for pairwise comparison, a method characterized by \cite{zheng2023judge}, who report over 80\% agreement with human preferences alongside position, verbosity, and self-enhancement biases. \cite{shi2025judging} show that position bias in capable judges is systematic rather than random, and varies across judges and tasks.
We mitigate position bias with randomized order over three runs and use the judge only for the large-scale ablation (Section~\ref{sec:results-ablation}), as a secondary instrument. These vulnerabilities, however, are not unique to machines: Chen et al.~\cite{chen2024humans} find that human judges likewise overlook injected factual errors and over-reward authority cues and presentation density, which provides support for our characterization of our human study as \textit{perceived} analysis quality.

%\textbf{Automated scientific discovery.}
%At the far extreme, The AI Scientist~\cite{lu2024aiscientist} generates ideas, runs experiments, and writes and reviews papers end-to-end. Its automated reviewer is necessarily shallower than ours, trading depth for lifecycle breadth.
%We aim not to replace the scientist but to produce a structured reading guide that makes their time more productive.
%\cite{thakur2025judging} find that only the largest judge models align well with human raters even on short-answer tasks, that judges tend toward leniency and are sensitive to prompt length and complexity, and that agreement alone overstates alignment.

%% file: sec_gauntlet.tex
\section{The Gauntlet Comprehension Pipeline}
\label{sec:pipeline}

\begin{figure}[t]
  \centering
  \includegraphics[width=\columnwidth]{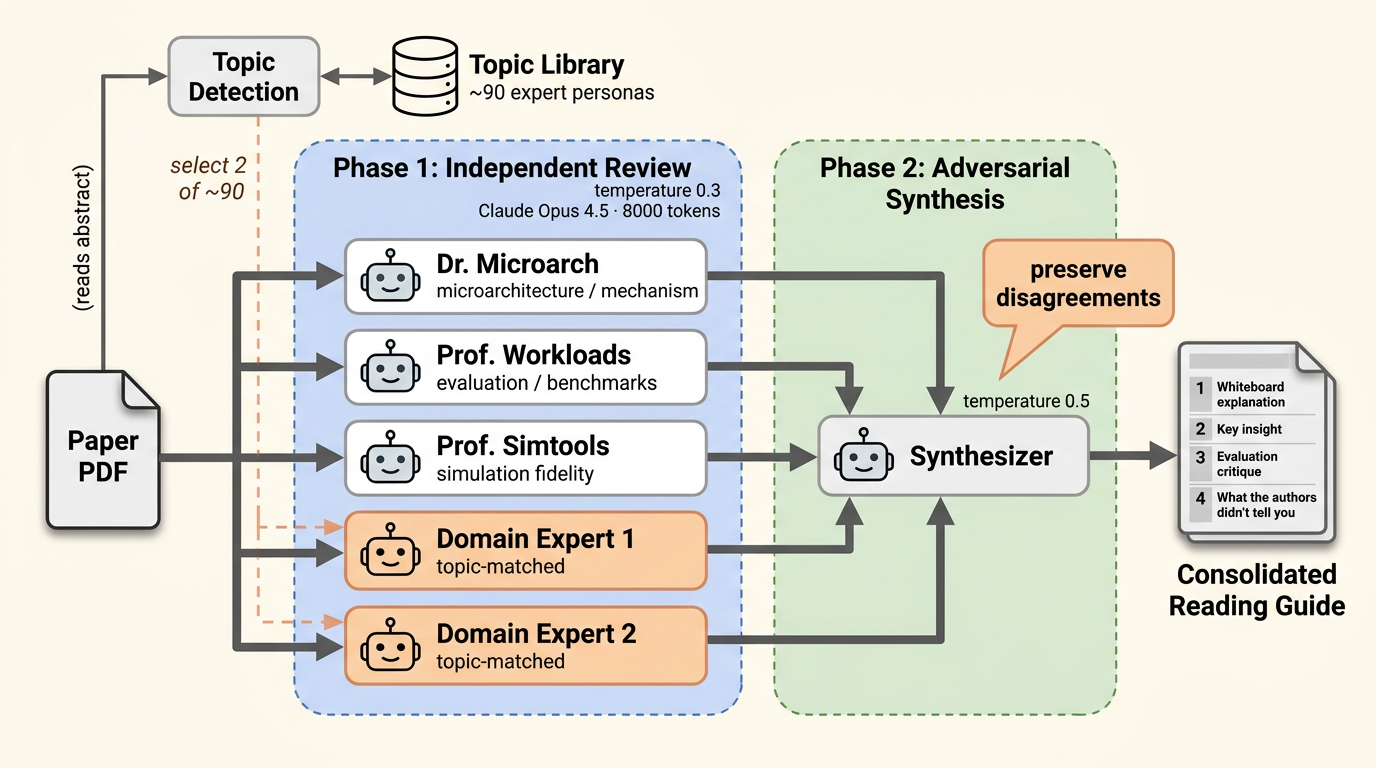}
  \caption{The Gauntlet paper-reading pipeline. Five reviewer agents analyze the paper independently---three fixed domain-general reviewers and two specialists dynamically selected from the paper's sub-topics. A synthesizer then produces a structured reading guide, explicitly prompted to surface disagreements between reviewers rather than average them away.}
  \label{fig:pipeline}
\end{figure}

Gauntlet reads a paper in two phases: independent multi-perspective review, then adversarial synthesis (Figure~\ref{fig:pipeline}).

\textbf{Phase 1: independent expert reviews.}
Five reviewer agents each analyze the paper over one iteration with no visibility into one another. Independence is deliberate, since shared context collapses distinct concerns into consensus and the most useful observations are smoothed away.
Three agents are fixed: \emph{Dr.\ Microarch} reverse-engineers the mechanism at the bit and structure level; \emph{Prof.\ Workloads} stress-tests the evaluation and benchmark selection; \emph{Prof.\ Simtools} audits simulation fidelity and reproducibility.
Two more are chosen per paper: an initial call identifies the paper's sub-topics and matches them against a library of $\sim$90 expert personas, instantiating the two closest---so a sparse-tensor paper draws a sparse-computation specialist rather than a generic reviewer.

\textbf{Phase 2: adversarial synthesis.}
A synthesizer consumes all five reviews plus the paper and produces a consolidated reading guide organized around the same four questions used for the human analyses (Section~\ref{sec:design}): the mechanism (a whiteboard explanation), the key insight (why it works), the evaluation critique, and what the authors did not tell you.
Its defining instruction is to \emph{preserve disagreement} rather than average it: where the microarchitect admires a design the workloads reviewer distrusts, the synthesis surfaces the tension instead of resolving it.

\textbf{Why the structure matters.}
No single reader---human or model---simultaneously holds the bit-level mechanism, the benchmark methodology, and the simulation fidelity in mind.
The pipeline instantiates these in parallel and extracts what becomes visible only in their comparison.
We validate this with a three-way ablation (Section~\ref{sec:results-ablation}): the ordering pipeline $>$ persona $>$ directive confirms the gain comes from structure, not prompt wording.

%% file: sec_expt_design.tex
\section{Experimental Design}
\label{sec:design}

\textbf{Corpus and analyses.}
Ten graduate-student volunteers each analyzed two computer-architecture papers in their area, drawn from ISCA 2025 and HPCA 2026 and restricted to concrete mechanism papers.
Each human analysis answered the same four questions Gauntlet does: mechanism, key insight, evaluation critique, hidden assumptions. We refer to the volunteers as \textit{analysts} in their role as paper evaluation writers. Each analyst had one week per paper and wrote about a paragraph per prompt. Analysts were forbidden from using LLMs to help comprehend papers or generate analyses. 
All Gauntlet analyses used Claude Opus 4.5, whose May 2025 knowledge cutoff predates the proceedings, making this comprehension rather than retrieval.

%Memory and caching: 'MagiCache', 'XOR Cache', 'MemSOS', and 'Memory Processing Unit'. Prefetching and prediction: 'Prophet', 'Rethinking Prefetching', and 'LLBP-X'. Accelerators: 'Qtenon', 'LightML', 'LUT Tensor Core', 'CoopRT', 'ELORA', and 'SLINFER'. Security and cryptography: 'Accelerating FHE'. Cores and systems: 'C3', 'Magellan', 'Forest', 'Hardware-aware Calibration Protocol', and 'Process Only Where You Look'. 

\textbf{Evaluation protocol.}
\label{sec:rubric}
For each paper we paired the human analysis with the Gauntlet synthesis.
The analysts judged papers they had not themselves analyzed, keeping the analyst and judge roles independent (in a few cases volunteers were unable due to time constraints to judge two papers; in these cases they were replaced by a senior reader). We refer to the volunteers and senior reader as \textit{judges} in their role as judging human vs. Gauntlet-written evaluations.
Judges were asked to read the original paper, or at minimum its abstract and introduction, and were given a week to complete both comparisons. They then scored both analyses on five dimensions on a 5-point scale: \emph{Mechanistic Accuracy} (is what was built described correctly?), \emph{Insight Depth} (the non-obvious why), \emph{Critical Rigor} (specific, genuine weaknesses), \emph{Calibration} (appropriately confident, not wrong at full confidence), and \emph{Usefulness} (would it prepare you for a meeting?).
A pilot sixth dimension, Breadth, was dropped as redundant.
Each judge also recorded an overall preference and a free-text justification.

Judges therefore scored from the same material a reader uses when triaging a paper. This supports assessment of specificity, clarity, and usefulness, but not independent verification of every mechanistic claim: a judge without direct expertise in a paper's subarea could assess whether a mechanism description was coherent and well-supported, not whether it was necessarily correct. We report the results accordingly as \textit{judged analysis quality}, and we treat Mechanistic Accuracy and Calibration as the dimensions most exposed to this limitation.

\textbf{Open-label evaluation.}
\label{sec:bias}
We intended to blind judges to which analysis was machine-generated, but a pilot showed Gauntlet's output trivially identifiable from its uniformity and completeness, and stylistic normalization did not hide it.
Rather than claim a blinding we could not achieve, we disclosed the source, which could bias judges in either direction.

\textbf{Ablation arm.}
\label{sec:ablation-design}
To separate the multi-agent structure from the model, we generated two single-agent baselines with the same model: \emph{Study~A} (a one-sentence prompt) and \emph{Study~B} (a rich, skeptical computer-architect persona); the full pipeline is \emph{Study~C}.
We ran all three over the full 98-paper corpus (80 ISCA 2025, 18 HPCA 2026; the 20 human-judged papers are a subset) and compared them pairwise with an automated judge---Gemini~3.1~Pro, blind, three randomized runs---reported separately as a secondary instrument (Section~\ref{sec:results-ablation})\footnote{The ablation uses a six-dimension variant of the rubric.}.

%% file: sec_results.tex
\section{Results}
\label{sec:results}

We treat each of the 20 comparisons as the unit of observation: each of the 20 papers appears once, with one human and one Gauntlet analysis, and one judge. Because judges scored one to four comparisons and analysts wrote two, comparisons are not fully independent. Therefore, we report confidence intervals clustered by judge and confirm the main results under tests which treat the judge as the independent unit and a mixed-effects model. The total-score advantage remains significant under every analysis, while Critical Rigor remains significant under all but judge-level averaging (Holm $p = 0.08$).

\begin{table}[t]
  \centering\small\setlength{\tabcolsep}{4pt}
  \begin{tabular}{lcccc}
    \toprule
    Dimension & Human & Gauntlet & $\Delta$ [95\% CI] & $p$ (Holm) \\
    MA & 3.80 & 4.60 & $+0.80$ [0.07,1.53] & $0.019$ \\
    ID       & 3.80 & 4.70 & $+0.90$ [0.09,1.71] & $0.017$ \\
    CR       & 3.65 & 4.75 & $+1.10$ [0.65,1.55] & $<0.001$ \\
    C          & 4.25 & 4.40 & $+0.15$ [-0.61,0.91] & $0.540$ \\
    U           & 3.65 & 4.60 & $+0.95$ [0.39,1.51] & $0.011$ \\
    \textbf{Total (/25)} & \textbf{19.15} & \textbf{23.05} & $\mathbf{+3.90}$ [0.95,6.85] & $\mathbf{<0.001}$ \\
    \bottomrule
  \end{tabular}
  \caption{Mean per-dimension and total scores (1--5 scale; total of 25), with mean delta and 95\% confidence intervals (CI), and two-sided Wilcoxon $p$-values with Holm correction (total score uncorrected). MA $=$ Mechanistic Accuracy, ID $=$ Insight Depth, CR $=$ Critical Rigor, C $=$ Calibration, U $=$ Usefulness}
  \vspace{-0.2in}
  \label{tab:perdim}
\end{table}

\textbf{Overall}
\label{sec:results-overall}
Evaluators preferred the Gauntlet synthesis in 15 of the 20 comparisons and the human analysis in 4, with one tie ($\sim79\%$ of decided comparisons). Gauntlet's mean total exceeded the human's by 3.9 points of 25. This result is robust to how dependence between comparisons is handled. It remains significant under tests which treats each judge as the independent unit ($p = 0.010$ under a permutation test, and $p=0.014$ under a per-judge average test), and in a mixed-effects model of total score with judge and analyst random effects in which judges are the largest source of variation ($+3.17$, 95\% CI [0.95, 5.39], $p = 0.005$).

\textbf{Per dimension.}
\label{sec:results-dims}
Two patterns are robust (Table~\ref{tab:perdim}).
\emph{Critical Rigor} is the strongest and most consistent signal: it gains $+1.1$ across all comparisons, the only dimension significant at $p<0.01$. It is also the only dimension that stays significant when the judge is treated as the independent unit. Evaluators credit Gauntlet's specificity, naming missing baselines, untested regimes, and buried assumptions where human critiques stay generic.
\emph{Calibration} is the only dimension where humans and Gauntlet are statistically indistinguishable ($p=0.54$, $\Delta = +0.15$). It is also where Gauntlet's characteristic failure surfaces. The other dimensions are suggestive rather than statistically significant.

\textbf{How Close are the Results?}
\label{sec:results-closeness}
In three of the four human wins, Gauntlet trailed by only 1--2 points. In the fourth (MagiCache) it led by 5 points but was not preferred due to a single wrong claim (Section~\ref{sec:results-calibration}). When Gauntlet won, its mean margin was by 5.13 points. 71\% of Gauntlet's dimension scores are 5 out of 5, and it scored a perfect 25 in 6 of 20 comparisons. By comparison, only 21\% of the human dimension scores are 5 out of 5, with no perfect 25's.

\if 0
\begin{figure}[t]
  \centering
  \fbox{\parbox{0.92\columnwidth}{\small \emph{Excerpt from Gauntlet's analysis of Qtenon, a quantum--classical accelerator (evaluation critique):}\\[3pt]
  ``The comparison uses an Ethernet-connected FPGA ($\sim$10\,ms latency, Table~1), but modern quantum systems (IBM, Google) use custom low-latency links, PCIe, or CXL at $\sim$100\,ns--1\,$\mu$s. A PCIe-attached baseline would shrink the reported $5000$--$6000\times$ communication speedups to $\sim$$100$--$1000\times$.''}}
  \caption{A representative excerpt of Gauntlet's output. The critique reasons from the paper's stated Ethernet--FPGA baseline to what deployed systems actually use, then quantifies the effect on the headline result---generative technical reasoning about this paper's specific mechanism and evaluation.}
  \label{fig:example}
\end{figure}
\fi

\subsection{Where Humans Still Win}
\label{sec:results-calibration}

The four human-preferred cases and the one tie isolate what the pipeline fails to do\footnote{We refer to these papers by the following shortened titles: 'MagiCache', 'Prophet', 'LLBP-X', 'LightML', and 'Qtenon'.}. We examined all five against the papers and the evaluators' justifications.
Even in the cases it \emph{lost}, Gauntlet keeps its analytic edge: it wins Critical Rigor in four of the five and Insight Depth in three. Human wins turn on whether an analysis can be \emph{trusted} and \emph{used}, not on depth. They fall into three modes.

\textbf{Trust: a confident, wrong claim.}
The only ``human clearly better'' verdict (MagiCache) turned on one sentence: Gauntlet asserted, at full confidence, a clearly erroneous interpretation of the paper's main contribution.
The evaluator scored Gauntlet higher overall (22 vs.\ 17) and called it the stronger review, yet preferred the human ``clearly.'' A single wrong claim, stated precisely and confidently in a weakness section, voided trust.
It was caught because the human reader had technical expertise.

\textbf{Teachability: mechanism told, not taught.}
For Prophet and LLBP-X, the human won narrowly (22--21 and 24--23), and in both
the judge scored the human higher on Mechanistic Accuracy (5 vs.\ 3 and 5
vs.\ 4) while Gauntlet led on Critical Rigor. Gauntlet's mechanism description was correct but not self-contained. For LLBP-X, it leaned on undefined shorthand and figure references, prompting the verdict that ``reading [the human review] is better than reading the paper.'' For Prophet, it gave a high-level gloss. The human instead walked through all policies and recomputed the paper's reported numbers.
A description the reader cannot rebuild the mechanism from fails the meeting-prep bar, which was scored as an accuracy deficit.

\textbf{Judgment: breadth without prioritization.}
For LightML, Gauntlet's weakness list was comprehensive but untriaged: it set trivial points (a breakdown of a particular result) beside load-bearing ones (an unvalidated simulator). The focused human review was preferred, and here over-coverage cost Gauntlet even Insight Depth and Critical Rigor.
The Qtenon tie shows a related \emph{decomposition bias}: Gauntlet detailed the hardware it could describe precisely but skipped the software contributions entirely, so it could not explain how the headline speedup arises. The human enumerated both types of contributions.

%These failure modes are the inverse of the pipeline's strengths. It maximizes coverage, but coverage alone does not guarantee a trustworthy claim, a teachable mechanism, or a prioritized account. These are properties of communication and judgment. The most damaging failure, the confident error, is a Calibration failure, the one dimension where humans hold parity.

\subsection{Ablation: Does the Multi-Agent Structure Matter?}
\label{sec:results-ablation}

\begin{table}[t]
  \centering\small
  \begin{tabular}{lccc}
    \toprule
    Comparison & Winner & Win rate & Mean $\Delta$ \\
    \midrule
    A vs.\ B (directive vs.\ persona) & B & (89\%) & $0.59$ \\
    A vs.\ C (directive vs.\ pipeline) & C & (99\%) & $1.04$ \\
    B vs.\ C (persona vs.\ pipeline)   & C & (96\%) & $0.58$ \\
    \bottomrule
  \end{tabular}
  \caption{Automated blind ablation over all 98 papers (Gemini~3.1~Pro, three runs; $\Delta$ is the score margin toward the winner, of 5). The ordering $C>B>A$ holds across both venues and every margin.}
  \vspace{-0.2in}
  \label{tab:ablation}
\end{table}

The ordering is unambiguous: $C>B>A$ (Table~\ref{tab:ablation}).
A rich persona beats a bare directive on 89\% of papers. The pipeline beats the directive near-unanimously (97/98) and, more tellingly, beats the \emph{strong persona} on 94 of 98 ($+0.58$).
The B$<$C result therefore isolates the Gauntlet structure as the source of the gain.
The effect is not a small-sample artifact: a 22-paper pilot showed the same ordering, and scaling to 98 papers improved the win rate from 73\% to 96\%.

Where C's edge narrows is instructive, and mirrors Section~\ref{sec:results-calibration}. The edge is \emph{contribution-shaped}: widest on broad, multi-mechanism papers ($\Delta = 1.60$ and $1.70$), and narrowest, sometimes negative, on single-trick papers. When there is one idea to extract, a focused single reading captures it, and the ensemble only adds dilution.

%% file: sec_discussion.tex
\section{Discussion and Data Release}
\label{sec:discussion}

\textbf{Data release.}
We release the pipeline, both rubrics, the 20 paired analyses with all scores
and justifications, and the A/B/C ablation reviews and judge transcripts for
all 98 papers---to our knowledge the first paired human--LLM dataset for
evaluating technical comprehension of architecture papers, all at \parallaxgitrepo.

\textbf{Limitations.}
Our evaluation measures judged analysis quality, not verified comprehension: judges scored from at least the abstract and introduction (the guaranteed floor), and MagiCache is the one error we know a judge's expertise caught. An objective factuality audit of the Gauntlet analyses by experts is the necessary next step. Our analysts are graduate students rather than senior researchers; they do most first-pass reading, so this is a realistic, but not the strongest, baseline. Judges and analysts mostly come from the same pool, so domain expertise varies across papers. The evaluation is also open-label (Section~\ref{sec:bias}). All analyses use one model and configuration; the transferable contribution is the architecture, not the model. Our ablation establishes that this structure beats a single agent or single-shot prompt, but does not disaggrgate the structure's components, which is left for future work.

\textbf{Use Cases.}
Gauntlet's analyses are preferred to human-written ones but can be confidently wrong, so it should not serve as a program-committee reviewer or a standalone teaching aid. Rather, it works as a companion the reader checks against the paper. For example, an expert can triage recent papers in their area, while catching errors like the MagiCache claim from their own knowledge. A student can write their own analysis first and then compare, to see whether they found the core insight and the load-bearing weaknesses, while verifying the claims Gauntlet flags.

\if 0
\textbf{A clarity phase before review.}\label{sec:clarity-phase}
The same machinery suggests a deployment at submission time, which we sketch as a proposal, not a validated system.
It follows from what we measured: the pipeline's strength is surfacing specific weaknesses (Critical Rigor) and its weakness is the occasional confident error (Calibration).
Enable authors to run their paper through the pipeline \emph{before} review, revise against the comprehension report over up to three iterations, and attach a $\leq$1000-word \emph{Clarification} flagging inaccuracies, submitted along with the final report; the pipeline never reports a score or verdict.
Reviewers then begin from the mechanism explanation, a curated weakness list, and the author's corrections.
We frame the benefits as hypotheses our findings motivate: authors see how a competent non-specialist will misread them before it costs a reviewer; reviewer effort shifts from comprehension toward judgment; weaknesses surface a cycle earlier.
The arrangement also turns the system's one weakness into an asset---the author, the maximal domain expert, is exactly who corrects a confident error (Section~\ref{sec:results-calibration}), and the word limit forces the prioritization the machine lacks.
Risks remain---reviewers may anchor on a wrong critique despite the Clarification---and measuring them is future work.
\fi

%% file: sec_appendix.tex
\appendix
\section{Case Studies: The Non-LLM-Win Analyses}
\label{sec:appendix-cases}

Of the 20 matched comparisons, Gauntlet was \emph{not} preferred in five: four human-preferred and one tie.
These five are the complete set---not a selection---and constitute the evidence for the failure modes in Section~\ref{sec:results-calibration}.
We summarize each below, with per-dimension scores as $[\text{Mechanistic Accuracy}, \text{Insight Depth}, \text{Critical Rigor}, \text{Calibration}, \text{Usefulness}]$ (Human~/~Gauntlet), and the decisive factor drawn from the evaluator's written justification and our reading of the paper.
Full deep dives are included in the data release.

\subsection*{Trust: a confident, specific, wrong claim}

\textbf{MagiCache} (fused computing-storage cache) --- \emph{Human clearly better}.
Human $[3,2,4,4,4]=17$; Gauntlet $[5,5,5,3,4]=22$.
Gauntlet was the higher-scoring review on four of five dimensions, and the evaluator called it ``likely the stronger review.''
It nonetheless lost ``clearly'' on one sentence: the claim that bit-line computation makes \emph{every} cache access pay a 60\% latency tax ($1.6$\,ns vs.\ $1.0$\,ns).
A per-line ``computing bit'' gates the slow path, so ordinary reads run at $1.0$\,ns; the dual-mode array is the paper's central point.
The confidently stated error voided trust in the analysis, and the evaluator---a cache domain expert---caught it immediately.

\subsection*{Teachability: mechanism told, not taught}

\textbf{Prophet} (profile-guided temporal prefetching) --- \emph{Human somewhat better}.
Human $[5,4,4,5,4]=22$; Gauntlet $[3,5,5,5,3]=21$.
Gauntlet led on Insight and Critical Rigor, but the human won on Mechanistic Accuracy and Usefulness by walking through the three interacting policies (insertion, replacement, resizing) at implementation resolution where Gauntlet stayed high-level.
The human also performed a domain-specific check Gauntlet missed: the paper's reported $1.6\%$ energy increase at a $35\%$ speedup implies a $\sim$$56\%$ increase in \emph{power}, exposing a misleading framing.

\textbf{LLBP-X} (last-level branch predictor) --- \emph{Human somewhat better}.
Human $[5,5,4,5,5]=24$; Gauntlet $[4,5,5,4,5]=23$.
The closest margin in the set.
Gauntlet found more weaknesses (a 40\% overprefetch rate, excluded Google traces, unquantified retraining), but its mechanism description leaned on undefined shorthand (``RCR,'' ``CID\_64'') and figure references, requiring the reader to consult the paper.
The human's plain-language, self-contained explanation drew the verdict that ``reading [it] is better than reading the paper itself.''

\subsection*{Judgment: breadth without prioritization or balance}

\textbf{LightML} (photonic ML accelerator) --- \emph{Human somewhat better}.
Human $[3,4,4,4,4]=19$; Gauntlet $[4,3,3,4,3]=17$.
Gauntlet was more complete on the mechanism but produced an untriaged weakness list, raising trivial points (an ADC power breakdown) alongside the load-bearing ones (an unvalidated simulator, no training-versus-inference distinction).
This is the one case where the coverage problem dragged down even Gauntlet's usual strengths, costing it Insight Depth and Critical Rigor; the human's focused review was preferred for concentrating on what matters.

\textbf{Qtenon} (hybrid quantum-classical architecture) --- \emph{Tie}.
Human $[4,3,3,4,4]=18$; Gauntlet $[3,4,4,4,3]=18$.
Gauntlet found the sharper criticism---that the headline $5000$--$6000\times$ communication speedup is measured against an Ethernet-attached FPGA baseline, and a PCIe/CXL baseline would shrink it by one to two orders of magnitude---and articulated quantum locality as the organizing principle.
But it analyzed only the hardware it could describe precisely and skipped the paper's software contributions (incremental compilation, the soft memory barrier, batched measurement) entirely, so it could not explain how the speedup actually arises.
The human enumerated all seven hardware and software contributions; the two reviews are genuinely complementary.

%% file: references.bib
@article{10.1109/LCA.2025.3559738,
author = {Yan, Liang and Lu, Xiaoyang and Chen, Xiaoming and Han, Yinhe and Sun, Xian-He},
title = {Pyramid: Accelerating LLM Inference With Cross-Level Processing-in-Memory},
year = {2025},
issue_date = {Jan.-June 2025},
publisher = {IEEE Computer Society},
address = {USA},
volume = {24},
number = {1},
issn = {1556-6056},
journal = {IEEE Comput. Archit. Lett.},
month = jan,
pages = {121–124},
numpages = {4}
}

@misc{gauntlet,
  title        = {Gauntlet},
  author       = {Sankaralingam, Karthikeyan},
  howpublished = {\url{https://github.com/VerticalResearchGroup/Gauntlet}},
  year         = {2025}
}

@misc{alphazeromoment,
  title        = {Computer Architecture's {AlphaZero} Moment: Automated Discovery in an Encircled World},
  author       = {Sankaralingam, Karthikeyan},
  howpublished = {arXiv:2604.03312},
  year         = {2026}
}

@misc{archagent,
  title        = {{ArchAgent}: Agentic {AI}-driven Computer Architecture Discovery},
  author       = {{Gupta~et~al.}},
  howpublished = {arXiv:2602.22425 [cs.AI]},
  year         = {2026}
}

@misc{quarch,
  title        = {{QuArch}: A Question-Answering Dataset for {AI} Agents in Computer Architecture. {IEEE CAL 2025.}},
  author={{Prakash~et~al.}}
}

@article{liang2024feedback,
  title={Can Large Language Models Provide Useful Feedback on Research Papers? A Large-Scale Empirical Analysis},
  author={Liang, Weixin and Zhang, Yuhui and Cao, Hancheng and Wang, Binglu and Ding, Daisy Yi and Yang, Xinyu and Vodrahalli, Kailas and He, Siyu and Smith, Daniel Scott and Yin, Yian and McFarland, Daniel A. and Zou, James},
  journal={NEJM AI},
  year={2024},
  publisher={Massachusetts Medical Society}
}

@inproceedings{liang2024monitoring,
  title={Monitoring {AI}-Modified Content at Scale: A Case Study on the Impact of {ChatGPT} on {AI} Conference Peer Reviews},
  author={Liang, Weixin and Izzo, Zachary and Zhang, Yaohui and Lepp, Haley and Cao, Hancheng and Zhao, Xuandong and Chen, Lingjiao and Ye, Haotian and Liu, Sheng and Huang, Zhi and McFarland, Daniel and Zou, James Y.},
  booktitle={ICML},
  year={2024}
}

@article{thakkar2025iclr,
  title={Can {LLM} Feedback Enhance Review Quality? A Randomized Study of {20K} Reviews at {ICLR} 2025},
  author={Thakkar, Nitya and Yuksekgonul, Mert and Silberg, Jake and Garg, Animesh and Peng, Nanyun and Sha, Fei and Yu, Rose and Vondrick, Carl and Zou, James},
  journal={Nature Machine Intelligence},
  year={2026},
  publisher={Springer Nature}
}

@inproceedings{du2023debate,
  title={Improving Factuality and Reasoning in Language Models through Multiagent Debate},
  author={Du, Yilun and Li, Shuang and Torralba, Antonio and Tenenbaum, Joshua B. and Mordatch, Igor},
  booktitle={ICML},
  pages={11837--11860},
  year={2024},
  volume={235},
  series={PMLR}
}

@inproceedings{wang2024spp,
  title={Unleashing the Emergent Cognitive Synergy in Large Language Models: A Task-Solving Agent through Multi-Persona Self-Collaboration},
  author={Wang, Zhenhailong and Mao, Shaoguang and Wu, Wenshan and Ge, Tao and Wei, Furu and Ji, Heng},
  booktitle={NAACL},
  year={2024},
  publisher={ACL}
}

@article{darcy2024marg,
  title={{MARG}: Multi-Agent Review Generation for Scientific Papers},
  author={D'Arcy, Mike and Downey, Doug and Hope, Tom},
  journal={arXiv preprint arXiv:2401.04259},
  year={2024}
}

@inproceedings{zheng2023judge,
  title={Judging {LLM}-as-a-Judge with {MT}-Bench and Chatbot Arena},
  author={Zheng, Lianmin and Chiang, Wei-Lin and Sheng, Ying and Zhuang, Siyuan and Wu, Zhanghao and Zhuang, Yonghao and Lin, Zi and Li, Zhuohan and Li, Dacheng and Xing, Eric P. and Zhang, Hao and Gonzalez, Joseph E. and Stoica, Ion},
  booktitle={NeurIPS},
  year={2023}
}

@article{demvsar2006statistical,
  title={Statistical comparisons of classifiers over multiple data sets},
  author={Dem{\v{s}}ar, Janez},
  journal={Journal of Machine learning research},
  volume={7},
  number={Jan},
  pages={1--30},
  year={2006}
}

@inproceedings{chen2024humans,
  title={Humans or LLMs as the judge? a study on judgement bias},
  author={Chen, Guiming Hardy and Chen, Shunian and Liu, Ziche and Jiang, Feng and Wang, Benyou},
  booktitle={Proceedings of the 2024 Conference on Empirical Methods in Natural Language Processing},
  pages={8301--8327},
  year={2024}
}

@inproceedings{van2019best,
  title={Best practices for the human evaluation of automatically generated text},
  author={Van Der Lee, Chris and Gatt, Albert and Van Miltenburg, Emiel and Wubben, Sander and Krahmer, Emiel},
  booktitle={Proceedings of the 12th international conference on natural language generation},
  pages={355--368},
  year={2019}
}

@inproceedings{shi2025judging,
  title={Judging the judges: A systematic study of position bias in llm-as-a-judge},
  author={Shi, Lin and Ma, Chiyu and Liang, Wenhua and Diao, Xingjian and Ma, Weicheng and Vosoughi, Soroush},
  booktitle={Proceedings of the 14th International Joint Conference on Natural Language Processing and the 4th Conference of the Asia-Pacific Chapter of the Association for Computational Linguistics},
  pages={292--314},
  year={2025}
}
